1-1-2010

# ON COURSE, BUT NOT THERE YET: ENTERPRISE ARCHITECTURE CONFORMANCE AND BENEFITS IN SYSTEMS DEVELOPMENT

Ralph Foorthuis
*UWV, Business Services*, ralph.foorthuis@uwv.nl

Marlies van Steenbergen
*Sogeti Nederland BV*, marlies.van.steenbergen@sogeti.nl

Nino Mushkudiani
*Statistics Netherlands*, n.mushkudiani@cbs.nl

Wiel Bruls
*IBM Netherlands NV*, wiel_bruls@nl.ibm.com

Sjaak Brinkkemper
*Utrecht University*, s.brinkkemper@cs.uu.nl

*See next page for additional authors*

**Authors**

Ralph Foorthuis, Marlies van Steenbergen, Nino Mushkudiani, Wiel Bruls, Sjaak Brinkkemper, and Rik Bos

# ON COURSE, BUT NOT THERE YET: ENTERPRISE ARCHITECTURE CONFORMANCE AND BENEFITS IN SYSTEMS DEVELOPMENT

*Completed Research Paper*

**Ralph Foorthuis**
UWV, Business Services
La Guardiaweg 116
1040 HG Amsterdam, The Netherlands
ralph.foorthuis@uwv.nl

**Marlies van Steenbergen**
Sogeti Nederland BV
Hofsemolenweg 5
8171 PM Vaassen, The Netherlands
marlies.van.steenbergen@sogeti.nl

**Nino Mushkudiani**
Statistics Netherlands
Henri Faasdreef 312
2492 JP The Hague, The Netherlands
n.mushkudiani@cbs.nl

**Wiel Bruls**
IBM Netherlands NV
David Ricardostraat 2-4
1066 JS Amsterdam, The Netherlands
wiel_bruls@nl.ibm.com

**Sjaak Brinkkemper**
Utrecht University, Institute of
Information and Computing Sciences
Padualaan 14
3584 CH Utrecht, The Netherlands
s.brinkkemper@cs.uu.nl

**Rik Bos**
Utrecht University, Institute of
Information and Computing Sciences
Padualaan 14
3584 CH Utrecht, The Netherlands
r.bos@cs.uu.nl

## Abstract

*Various claims have been made regarding the benefits that Enterprise Architecture (EA) delivers for both individual systems development projects and the organization as a whole. This paper presents the statistical findings of a survey study (n=293) carried out to empirically test these claims. First, we investigated which techniques are used in practice to stimulate conformance to EA. Secondly, we studied which benefits are actually gained. Thirdly, we verified whether EA creators (e.g. enterprise architects) and EA users (e.g. project members) differ in their perceptions regarding EA. Finally, we investigated which of the applied techniques most effectively increase project conformance to and effectiveness of EA. A multivariate regression analysis demonstrates that three techniques have a major impact on conformance: carrying out compliance assessments, management propagation of EA and providing assistance to projects. Although project conformance plays a central role in reaping various benefits at both the organizational and the project level, it is shown that a number of important benefits have not yet been fully achieved.*

## Introduction

By providing holistic overviews and high-level constraints, guidelines and logic, Enterprise Architecture (EA) aims to achieve coherent and goal-oriented organizational processes, structures, information provision and technology (cf. Boh and Yellin 2007; Richardson, Jackson and Dickson 1990; Ross, Weill and Robertson 2006; The Open Group 2009; Wagter, Van den Berg, Luijpers and Van Steenbergen 2005). Various claims have been made regarding the application and effectiveness of EA, by academics and practitioners alike. At the level of the *entire organization*, for example, benefits in the reduction of complexity and realization of business/IT alignment have been claimed. At the level of *individual systems development projects*, costs and risks are said to be brought down when complying with EA. Since conformance to EA is crucial for gaining the aforementioned benefits (Boh and Yellin 2007; Foorthuis et al. 2009; Goodhue et al. 1992), various techniques for stimulating project compliance with EA are suggested in the literature. A complete overview of techniques and benefits claimed by academics and practitioners will be provided in a subsequent section of this paper.

As the need for hypothesis testing on the topic of Enterprise Architecture has been identified in the IS research community (Boh and Yellin 2007; Kappelman et al. 2008; Niemi 2006), this paper aims to critically and empirically verify various claims regarding EA benefits and conformance. To add to theory building, we constructed an empirically supported regression model of determinants of EA success, mainly in the context of projects. An important element here is the extent to which projects conform to the EA's norms or prescriptions (i.e. principles, rules, standards, guidelines, models, et cetera). Both EA creators (e.g. enterprise architects and managers responsible for delivering the Enterprise Architecture) and EA users (e.g. project members applying EA constraints) contributed important perspectives to our research. The first stakeholder group brought in important perspectives regarding enterprise-wide aspects that remain invisible to local appliers. The latter group offered views regarding actually applying EA in practice, aspects of which might remain invisible to the creators of architectural constraints and guidelines. Since we investigated the perceptions of both groups, it was also relevant to test whether differences in evaluations of EA exist.

In short, the high-level research question of this paper is:

> *What benefits can be gained by conforming to EA, and what are the most effective techniques for achieving conformance?*

To answer the main research question, several sub-questions need to be taken into account:

1. What techniques are applied in practice to stimulate conformance of projects to EA?
2. What benefits does EA yield for individual projects conforming to its norms?
3. What benefits does EA yield for the enterprise as a whole?
4. What differences, if any, are there between EA users and EA creators in their evaluative perceptions?
5. Which of the techniques applied result in the most effective increase in conformance to EA?

The overall theoretical framework is presented in the next section. Following this, the specific claims regarding techniques and benefits – serving as the hypotheses to be tested – are described in more detail by means of a literature review. The subsequent sections present the empirical research approach and results respectively. The final section describes the conclusions.

## Concepts in Enterprise Architecture Conformance and Benefits

This section discusses how the concepts in our study of the application and effectiveness of EA are interrelated. Several techniques can be used to stimulate project conformance to EA. Amongst others, for example, assistance can be offered to projects when applying EA prescriptions or projects can be assessed on compliance. See the next section for a complete overview of techniques. Employment of these techniques should lead to conformance of projects to EA, which, in turn, should result in reaping the aforementioned EA-related benefits. The claimed benefits of conformance to EA are multifold. For the *organization* as a whole, for example, a coherent enterprise-wide strategy can be implemented – instead of local optimizations – and business and IT can be aligned. For *projects*, for example, costs, duration, complexity and risk are said to be reduced. The next section also presents a complete overview of benefits.

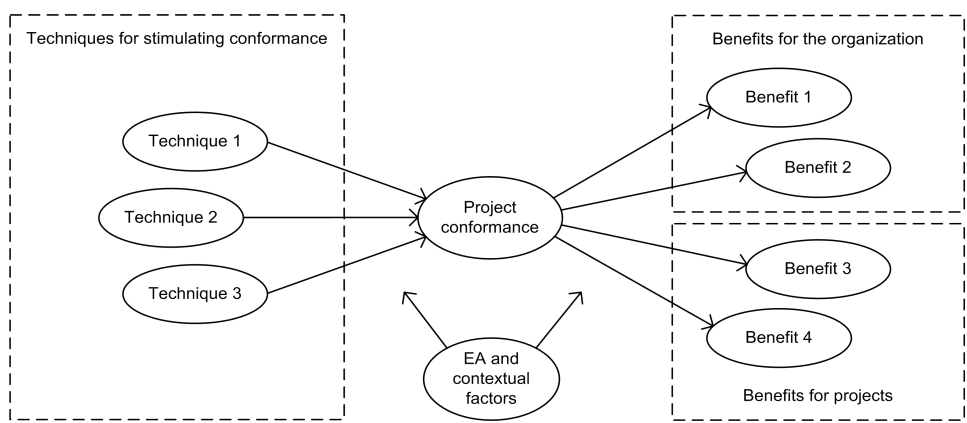

**Figure 1. Theoretical framework for project conformance and EA benefits**

Furthermore, the effects of techniques and conformance might be influenced by several contextual factors, such as the economic sector, organizational size and EA focus (on business, IT or both). The diagram above visually presents the overall theoretical framework, which has been used to structure our research and to construct an empirically supported model for EA conformance and benefits.

Needless to say, in addition to conformance-stimulating techniques, more generic techniques are equally crucial for success (Ward and Peppard 2002). Examples of such best practices are the use of proven project management and system development approaches, and involving high-quality staff in projects. Furthermore, all this takes place in a specific context, where organizational culture, leadership and market conditions have an effect on the organization and its projects. However, these techniques are less relevant in this study since we focus on EA-related aspects.

## Overview of Claimed Techniques and Benefits

We will now describe in more detail the techniques for encouraging EA conformance and the benefits that can be gained when EA is actually applied. These techniques and benefits are drawn from both academic and practitioner publications and will serve as the hypotheses to be tested.

### Techniques for Stimulating Conformance to Enterprise Architecture

To be able to reap its benefits, it is important that an EA is actually complied with (Boh and Yellin 2007; Foorthuis et al. 2009; Goodhue et al. 1992). This sub-section provides an overview of techniques that can be employed to increase conformance to EA. For later reference, each technique is coded in parentheses with a capital T (e.g. T1).

**Ensure management involvement in EA**. EA should enable the achievement of strategic business goals (Morganwalp and Sage 2004; Obitz and Babu K 2009). In this context, it may be important for management to formally approve the EA (T1) (Van Steenbergen et al. 2010). Management should also ensure that the choices in the EA are explicitly linked to the strategic business goals (T2). Furthermore, it is necessary for management to actively propagate the importance of EA for achieving these goals (T3) (Boh and Yellin 2007).

**Assess EA conformance.** Monitor projects and other initiatives on compliance with the EA's constraints and standards (T4), and use the results to take corrective action (Boh and Yellin 2007; Ellis 1993; Foorthuis et al. 2009).

**Create an active community for EA knowledge exchange.** The division of architectural domains over a number of domain architects, which is often felt necessary in large organizations, carries the inherent risk of fragmentation and misalignment. An active community of EA practice should enable knowledge integration (Van Steenbergen and Brinkkemper 2009). This manifests itself in organized knowledge exchanges between architects themselves (T5) and between architects and project members (T6). Moreover, some authors stress that architects should be actively involved in projects, as they can assist the projects in defining the solution and applying EA norms (T7). For example, architects could be used as consultants to periodically provide advice, or they can actively participate in the project or some of its stages (Foorthuis et al. 2008, 2009; Slot et al. 2009; Wagter et al. 2006).

**Leverage the value of project artifacts.** A Project Start Architecture (PSA) is a document created at the beginning of a project. This deliverable inherits and translates the EA's prescriptions – such as rules, guidelines and models – to the specific project situation (Wagter et al. 2006). In this context, a PSA can be regarded as a specific form of what in TOGAF (The Open Group Architecture Framework) is referred to as an architecture contract (The Open Group 2009). It describes the tangible constraints within which the project must operate. Using a PSA (T8) can therefore stimulate the project to comply with the EA's norms (Wagter et al. 2006). In fact, document templates in general can be employed to encourage conformance (T9) for several reasons. First, they seem likely candidates to be used as 'boundary artifacts' to encourage knowledge integration between architects, both between organizational levels and between domains (Van Steenbergen and Brinkkemper 2009). Secondly, templates can be used to provide projects with pre-defined structures and content prescribed by the EA, and provide the authors with instructions on how to conform (Foorthuis and Brinkkemper 2008; Wagter 2006).

**Use compensation or sanctioning for stimulating conformance.** Incentives and disincentives could increase the readiness of projects to comply with EA (T10). For example, the IT-costs of conformance might be compensated for by the EA program (cf. Foorthuis et al. 2009).

## Claimed Benefits of Working with Enterprise Architecture

We will now describe the benefits, which shall be indicated with a capital B (e.g. B1). The benefits of working with EA can be identified at two levels. First, the *organization as a whole* should profit from EA in several ways:

**EA enables management to achieve key business goals.** First, EA is said to enable management to pursue a coherent strategy that is optimal for the entire enterprise, instead of local optimizations (B1). Individual domains and departments may strive to pursue local interests. However, the firm as a whole will not benefit from conflicting goals and an EA can provide the required holistic view of the enterprise to balance different interests and solutions (Lankhorst et al. 2005). In addition, by taking a holistic and multi-layered view, Enterprise Architecture is a valuable instrument in aligning IT and the business processes it supports (B2) (Bucher et al. 2006; Gregor et al. 2007; Lankhorst et al. 2005). This is crucial, as business/IT alignment is an important instrument in realizing organizational value from IT investments (Henderson and Venkatraman 1993).

**EA enables management of organizational complexity.** Architecture can provide insight into complex problems (B3). Insight can be gained by means of different aspect areas (e.g. business, information, information systems and infrastructure) and levels of abstraction (Capgemini 2007; Raadt et al. 2004). Complexity can be managed (B4) by using a modular approach – which distinguishes between parts of a system and their relationships – and architectural modeling languages (Lankhorst et al. 2005; Versteeg and Bouwman 2006). Furthermore, implementing standardized and automated processes should result in less complex technology environments (Ross et al. 2006).

**EA facilitates the integration, standardization and deduplication of processes and systems.** Years of organic growth have often led to various 'silos' or 'stovepipes', which do not leverage the potential of related or similar processes and systems. The high-level overviews of an EA provide insights into the organization's processes and business and information systems. This enables the enterprise to identify processes that could be integrated (since it is beneficial to share valuable information), standardized (since similar processes can be supported by the same systems) or even cut out (since redundancy can be replaced by similar processes) (B5) (Capgemini 2007; Morganwalp and Sage 2004; Niemi 2006; Ross et al. 2006). As a result, costs can be controlled (B6).

**EA enables the enterprise to deal with its environment effectively.** Markets and businesses change ever more rapidly nowadays, and business processes and systems are highly interdependent. This poses IT problems, as software has to be updated or replaced sooner whilst simultaneously being part of an increasingly complex network of processes and systems. The agility of the enterprise's reaction to the outside world can be improved by EA (B7) by automating the core business processes. This results not only in more resources, but also in valuable information, which can be utilized in innovative activities (Ross et al. 2006). In addition, by focusing on the contextual relationships, an EA can facilitate co-operation with other organizations (B8) (Jonkers et al. 2006; Morganwalp and Sage 2004).

**EA enables effective communication between members of the organization.** EA provides a consistent and coherent overview of the fundamental aspects of the organization and the desired future situation (B9). This includes defining and interrelating concepts, e.g. by using models. EA thus provides members of the organization with a shared frame of reference to communicate effectively with each other (B10) (Armour et al. 1999; Bernus 2003; Foorthuis et al. 2009; Kappelman 2008; Raadt et al. 2004).

In addition to benefits for the organization, *individual projects* should also benefit from conforming to EA:

**Working with EA reduces project costs and project duration.** Projects can be expected to save resources (B12) and time (B13) when working in the context of EA, since its business, information, applications and technology decisions guide development work. EA and domain level decisions are a given starting point and do not have to be discussed inside the project (Capgemini 2007; Mulholland and Macaulay 2006; Pulkkinen and Hirvonen 2005; Wagter et al. 2005). A project can thus quickly focus its attention on designing and developing the details of the solution.

**Working with EA reduces project risk and improves project success.** Although publications tend to discuss the topic only superficially, EA is said to identify and mitigate project risks (B14). The argument that usually is put forward is that EA models – with their views on platforms, applications, processes and connections to other projects – provide insight into project risks (Bucher 2006; Capgemini 2007), allowing for timely risk prevention tactics. In addition, projects that conform to EA can benefit from the fact that issues at the enterprise-level have already been solved in the EA, thus mitigating risk and improving the chances of success, instead of building on sand (Capgemini 2007; Mulholland and Macaulay 2006; Pulkkinen and Hirvonen 2005). On a similar note, EA can be used to align the project with its context, resulting in high quality (B15) and relevant functionality (B16).

**Working with EA enables projects to manage complexity.** EA is said to enable projects to deal with complexity (B17) (Capgemini 2007). Analogous to controlling complexity at the organizational level, EA facilitates management of project complexity by using aspect areas, levels of abstraction, a modular approach, up-front decision making, and by standardized services, processes and systems (ibid.; Lankhorst et al. 2005; Ross et al. 2006). This should simplify project tasks, especially since certain issues should already have been resolved by the EA.

**Working with EA speeds up the initialization of a project.** An EA provides models of the enterprise, which help to specify the project scope and avoid redundant development activities (Bucher et al. 2006). Furthermore, several decisions have been made up-front and can be readily leveraged, e.g. by using a PSA (Wagter et al. 2005). Therefore, projects that have to conform to EA are expected to get initialized relatively fast (B18).

## Research Design

The research method for testing the hypotheses and model is an online survey and subsequent statistical analysis of its perceptual measures. The target population is defined as "all people working in the Netherlands in commercial or public organizations (either as internal employees or external consultants) who in a professional capacity have to deal with Enterprise Architecture (either as an EA creator, an EA user or both)." The unit of analysis is the individual worker, who is asked about his or her perceptions of EA functioning in practice. The reasons for choosing this unit of analysis is that it allows for obtaining information from different perspectives (enterprise architects, managers, system analysts, software architects, et cetera) and for asking questions referring to the project and organization level alike (from an individual's perspective, that is). When choosing e.g. a project as the unit of analysis, it is more difficult to obtain these divergent views (as the questionnaire is usually filled in only by the project manager involved) and levels (as analysis is bound to the project as object of study).

The design of the survey underwent several iterations. The questionnaire was first created as a written document by the principal researcher, which was reviewed by the other authors and a questionnaire expert from Statistics Netherlands. This led to several improvements in the design, e.g. using question-dependent response categories and textual clarifications. The resulting questionnaire design was used to create the web-survey, which was similarly reviewed. Finally, we verified the web-survey with three test-respondents (who worked for a government agency and an IT service provider). They had to fill in the questionnaire whilst articulating their thoughts out loud, so as to let us gain insight into the minds of respondents. This yielded several small simplifications and clarifications.

Since we did not specifically target executives, promising benchmarking reports would not have increased responses. We therefore opted for a relatively short questionnaire, containing 45 questions. At any moment during the survey session, respondents could see what percentage of the questions they had filled in, encouraging them to complete it. Because of the relatively short questionnaire, we had limited opportunities to create complex aggregated constructs.

All survey questions explicitly referred to the current (or latest) organization in which the respondents actually carried out their work, for example because they were an employee or because they were posted there as an external consultant. The first survey question explicitly asked whether the respondent has to deal with EA in his or her work (if not, the survey was finished). In general, questions featured five closed response categories (e.g. from *Very poor* to *Very good*) and one *No answer* category (in case the respondent did not know or did not want to provide an answer).

Since registers containing contact information of the individuals comprising our target population were not available, we used several communities related to information systems and architecture. First, an e-mail containing the hyperlink to the web-survey was sent to relations and employees of several IT-service providers and IT-intensive organizations. Secondly, the web-survey was advertised at two architecture conferences in the Netherlands attended mainly by practitioners. The data were gathered between October 2009 and May 2010.

In total, we received 293 valid surveys. A questionnaire had to pass several checks to be accepted as valid. First, it had to be completed and submitted. Secondly, it was checked whether there was a basic consistency between key variables. If the respondent's score on B11 was (very) poor, then B1, B2 and B7 were not allowed to all be scored (very) good. Likewise, if the respondent's score on B11 was (very) good, then B1, B2 and B7 could not be (very) poor. Thirdly, since we did not work with website passwords, we performed basic duplicate records checks. However, only unique records were found – and it required merely 11 included variables for a query to return zero duplicates.

## Research Results

The following four sub-sections present: descriptive statistics and sample representativeness; results of testing simple (singular) hypotheses; differences between EA creators and EA users; regression analysis and the resulting model.

### Descriptives and Representativeness

The respondents were working for 116 different organizations. Whereas general questions yielded high responses, several individual questions resulted in quite some nonresponse (i.e. item-nonresponse). This was probably due to the fact that such questions demanded very specific experience, which not every respondent possessed. Males dominate the field, 268 in total (91.5%), versus only 21 (7.2%) of the respondents being female and 4 (1.4%) unspecified. The tables below present the distribution of organizational roles. With regard to Table 1, note that individuals can work in multiple roles, resulting in a total of over 100%. Because of its relevance to our research goals, we consider it desirable to have a sample distribution that features a roughly equal number of EA users and EA creators. We conclude from Table 2 that this condition is satisfied.

**Table 1. Roles occupied by respondents (multiple roles allowed)**

| Role | | Frequency | Percentage |
|------|------|------|------|
| EA Creator | Enterpr Architect Business & Inform | 97 | 33.1% |
| | Enterpr Architect Application & Infrastr | 95 | 32.4% |
| | Manager | 39 | 13.3% |
| | External EA Consultant | 19 | 6.5% |
| EA User | Manager | 42 | 14.3% |
| | Project Manager | 39 | 13.3% |
| | Project Architect | 56 | 19.1% |
| | Business Analyst/Designer | 34 | 11.6% |
| | System & Information Analyst/Functional Designer | 26 | 8.9% |
| | Software Architect | 35 | 11.9% |
| | Technical Designer | 19 | 6.5% |
| | Developer/Programmer | 8 | 2.7% |
| | Maintenance Engineer | 8 | 2.7% |

**Table 2. EA creators and users**

| Role | Frequency | Percentage |
|------|------|------|
| EA Creator | 107 | 36.5% |
| EA User | 109 | 37.2% |
| EA Creator and User | 65 | 22.2% |
| Unknown | 12 | 4.1% |
| Total | 293 | 100% |

**Table 3. Organizational size**

| #Employees | Frequency | Percentage |
|------|------|------|
| <2000 | 81 | 27.7% |
| 2000-5000 | 78 | 26.6% |
| ≥5000 | 128 | 43.7% |
| Unknown | 6 | 2.0% |
| Total | 293 | 100% |

In order to assess the representativeness of our sample, we looked at the economic sectors. Since we could not use a pre-defined sampling frame corresponding with our target population (e.g. a population register of EA stakeholders), it was difficult to determine the extent to which our sample constitutes a representative subset. However, we could use previous research on EA for a comparison of sector distributions. Table 4 presents the distributions of our survey and those of two other studies. The column on the right (in white) presents the distribution found by Obitz & Babu K (2009) in their survey with 173 respondents drawn from the global IT community (mainly North America). The middle column (in grey) contains our results. The left column presents the distribution found by Bucher et al. (2006).

| Table 4. Distribution of respondents across economic sectors | | | | | | |
|---|---|---|---|---|---|---|
| **Bucher et al. (2006)** | | **Research in this paper** | | | **Obitz and Babu K (2009)** | |
| **Industry** | **%** | **Industry (based on ISIC Rev. 4)** | **%** | | **%** | **Industry** |
| Finance and insurance | 62.5 | 61.8 | Financial and insurance activities | 30.4 | 30.4 | 28.4 | Banks, insurance, financial services and capital markets |
| Industry | 12.5 | 13.2 | Manufacturing (products and food) and construction | 5.5 | 6.5 | 4.9 | Industrial goods/engineering, food, beverage, tobacco, pharmaceuticals & biotech |
| | | | Agriculture, fishing, forestry and mining | 1.0 | | | |
| Software, IT and Telecommunication | 25 | 25.0 | Information, communication, entertainment and recreation | 12.3 | 12.3 | 16.1 | Media, information, entertainment, telecom services, travel and leisure |
| Total | 100% | | Public administration (including defense) | 31.1 | 32.8 | 23.5 | Government, education, aerospace and defense |
| | | | Education and research | 1.7 | | | |
| | | | Energy & water supply and waste management | 5.1 | 5.1 | 6.2 | Oil, gas and utilities |
| | | | Human health and social work activities | 2.7 | 2.7 | 7.3 | Healthcare services |
| | | | Trade, transportation, hotel, catering, real estate and other services | 10.2 | 10.2 | 13.6 | Retail, transportation, logistics and other |
| | | | Unknown | 0 | 0 | N.a. | Professional services |
| | | | Total | 100% | 100% | 100% | Total |

Since we asked consultants posted at clients' offices to fill in the questionnaire from the perspective of their customer organization, this group was largely dispersed amongst the other sectors. We therefore have also dispersed the *Professional services* category of Obitz and Babu K (2009) in order to obtain a better comparison. The left and right columns demonstrate that the various industry distributions of these two studies are fairly similar. Our collapsed *Public administration, education and research* category is large, however, compared to Obitz & Babu K. This is consistent with the fact that the public sector in the Netherlands is large compared to North America. The left column presents the distribution of Bucher et al. (2006) of the respondents who indicated that EA is "in use". It features only a subset of our economic sectors, but this distribution is also similar to that of our sample. All in all, given the fact that the distributions are largely similar, we assume there is no reason to suspect that our distribution of economic sectors is not representative.

## Testing Simple Hypotheses on Conformance and Benefits

This sub-section will answer the first three sub-questions of this paper. Since we had no prior theoretical or empirical information about the distribution of the scores, we used a non-parametrical test. Each statement was subjected to a one-sided *binomial* test. The test proportion used was dependent on the type of question. For example, a first type of question, factual by nature, verifies specific characteristics of the EA approach (e.g. whether projects are being assessed on compliance with EA). The null hypothesis, assuming the distribution $B(n, 0.4)$, states that EA approaches generally do not possess the respective characteristic in abundance and is thus supported by the *Never* and *Seldom* response categories (one would consequently expect 40% or more of the respondents to fall in either of these two categories). The alternative hypothesis, which states that less than 40% falls within these two categories, is therefore supported by the *Sometimes*, *Frequently* and *Always* categories (i.e. assuming that EA approaches generally do possess the respective characteristic, one would expect significantly more than 60% of the answers to fall in these three categories). See Table 5 (except T1) for examples of this type of question. A second type of question, evaluative by nature, verifies whether EA is seen as a valuable instrument for achieving a specific goal (such as cost reduction). Here, the null hypothesis, assuming the distribution $B(n, 0.6)$, states EA is not particularly valuable and is supported by the *Very poor*, *Poor* and *Neither good nor poor* response categories (i.e. at least 60% is expected to fall in either of these three categories) and the alternative hypothesis covers the *Very good* and *Good* categories (i.e. more than 40% is expected). See Table 6 for examples of this type of question. Regardless of the type, respondents in the *No answer* category were always treated as missing values and therefore not included in the test.

The first column of the table below contains the statement from the survey (i.e. the alternative hypothesis representing the claim to be tested), the second to sixth columns (except for T1) provide the valid percentage of answers, the seventh gives the number of nonrespondents (which are invalid and therefore excluded from the binomial test), the eighth presents the p-value and the final column indicates whether the alternative hypothesis was accepted (p-values between 0.05 and 0.10 are seen as inconclusive). The grey columns present the percentages of respondents that support the alternative hypothesis. An underlined percentage denotes the respective value as the median.

One-sided tests were performed. An asterisk in the p-value column therefore indicates that a reversed test needed to be performed for that specific claim. That is, unlike the other hypotheses, the low p-value of this hypothesis shows that significantly more than 40% of the respondents fall in the *Never* or *Seldom* categories. A similar logic applies to statements in other tables with a significant p-value but a rejected H1 hypothesis.

| Table 5. Techniques used for stimulating compliance | | | | | |
|---|---|---|---|---|---|
| **H1 statement** | **No** | **Yes** | **NR** | **P-value** | **H1?** |
| T1. The EA is formally approved (i.e. by line management). | 20.4% | 79.6% | 19 | 0.000 | ✔ |

| **H1 statement** | **Never** | **Seldom** | **Some-times** | **Frequ-ently** | **Always** | **NR** | **P-value** | **H1?** |
|---|---|---|---|---|---|---|---|---|
| T2. The choices made in the EA are explicitly linked to the business goals of the enterprise as a whole. | 0.4% | 4.9% | 28.5% | 52.5% | 13.7% | 9 | 0.000 | ✔ |
| T3. Management propagates the importance of EA. | 2.8% | 19.3% | 46.4% | 27.3% | 4.2% | 4 | 0.000 | ✔ |
| T4. Projects are being explicitly assessed on their degree of compliance with EA. [Note: this concerns the number of projects being judged on compliance (not the number of times one project is being assessed).] | 4.8% | 11.1% | 24.9% | 42.9% | 16.3% | 4 | 0.000 | ✔ |
| T5. There is an organized knowledge exchange between different types of architects (for example enterprise, domain, project, software and infrastructure architects). | 1.5% | 12.9% | 34.3% | 44.3% | 7.0% | 22 | 0.000 | ✔ |
| T6. There is an organized knowledge exchange between architects and other employees participating in projects that have to conform to EA (for example project managers, functional designers, developers and testers). | 2.5% | 15.4% | 44.2% | 35.4% | 2.5% | 8 | 0.000 | ✔ |
| T7. Assistance is being offered in order to stimulate conformance to EA. (For example enterprise architects or change managers who help projects to make new designs conform to EA.) | 4.9% | 19.3% | 34.3% | 36.1% | 5.4% | 13 | 0.000 | ✔ |
| T8. Projects make use of a PSA (Project Start Architecture). | 8.9% | 10.0% | 24.4% | 40.9% | 15.8% | 14 | 0.000 | ✔ |
| T9. Document templates are being used to stimulate conformance to EA. (For example templates that focus attention on the EA by means of guiding texts and by requiring filling in relevant information.) | 5.0% | 11.4% | 33.4% | 37.0% | 13.2% | 12 | 0.000 | ✔ |
| T10. Financial rewards and disincentives are being used in order to stimulate conformance to EA. (For example covering the IT-expenses of a project if the solution is designed and built conform EA, or by imposing a fine for non-conformance.) | 65.6% | 17.9% | 10.3% | 5.5% | 0.7% | 20 | 0.000* | ✘ |

In general, we can conclude that most techniques for encouraging compliance are used regularly, as indicated by the acceptance of the alternative hypotheses in almost all cases. However, quite notable is the fact that financial sanctions are not being used to stimulate EA compliance. No less than 83.5% of the respondents indicated that they are *Seldom* or *Never* used, with the median even falling within the most extreme category (i.e. the lowest level). It can be hypothesized that financial penalties are difficult to implement (perhaps especially so in the Dutch culture of tolerance and compromise). However, this survey question also referred to financial incentives and presented respondents with an example of providing projects with free IT-resources if they should conform to EA. It is possible that financial sanctioning (rewarding or punishing) is a tactic that has unexploited potential.

Tables 6, 7 and 8 present the evaluations of the benefits of EA. In general, extreme scores, such as *Very good* or *Very poor*, were not commonly given by the respondents. This is not entirely unexpected, since the respondents are professionals reporting about their work and the subject matter is not as sensitive as certain social or political issues. In addition, most respondents only had a few years of experience with EA, possibly resulting in moderate attitudes.

| Table 6. Evaluations of EA benefits for the organization as a whole | | | | | | | | |
|---|---|---|---|---|---|---|---|---|
| **H1 statement**<br>EA turns out to be a good instrument to… | **Very poor** | **Poor** | **Neither good nor poor** | **Good** | **Very good** | **NR** | **P-value** | **H1?** |
| B1. …accomplish enterprise-wide goals, instead of (possibly conflicting) local optimizations. | 2.5% | 13.5% | 31.3% | <u>44.3%</u> | 8.4% | 18 | 0.000 | ✔ |
| B2. …achieve an optimal fit between IT and the business processes it supports. | 2.6% | 16.6% | <u>38.0%</u> | 40.6% | 2.2% | 22 | 0.189 | ✘ |
| B3. …provide insight into the complexity of the organization. | 1.1% | 6.4% | 18.4% | <u>64.6%</u> | 9.5% | 10 | 0.000 | ✔ |
| B4. …control the complexity of the organization. | 1.9% | 21.1% | <u>47.6%</u> | 27.6% | 1.8% | 18 | 0.000* | ✘ |
| B5. …integrate, standardize and/or deduplicate related processes and systems. | 2.2% | 11.2% | 31.0% | <u>48.4%</u> | 7.2% | 16 | 0.000* | ✔ |
| B6. …control costs. | 5.4% | 31.8% | <u>49.4%</u> | 11.5% | 1.9% | 32 | 0.000* | ✘ |
| B7. …enable the organization to respond to changes in the outside world in an agile fashion. | 1.9% | 22.6% | <u>50.2%</u> | 24.2% | 1.1% | 28 | 0.000* | ✘ |
| B8. …co-operate with other organizations effectively and efficiently. | 1.6% | 14.3% | <u>55.9%</u> | 25.8% | 2.4% | 48 | 0.000* | ✘ |
| B9. …depict a clear image of the desired future situation. | 0.4% | 4.2% | 23.5% | <u>59.3%</u> | 12.6% | 8 | 0.000 | ✔ |
| B10. EA turns out to be a good frame of reference to enable different stakeholders to communicate with each other effectively. | 0.7% | 13.0% | <u>40.1%</u> | 43.0% | 3.2% | 16 | 0.021 | ✔ |
| B11. EA, in general, turns out to be a good instrument. | 1.1% | 7.9% | 30.0% | <u>56.7%</u> | 4.3% | 13 | 0.000 | ✔ |

Looking at Table 6, we see that EA, in general, is considered to be a good instrument (B11). Several notable findings are worth looking into. Although several individual hypotheses are accepted, it seems that most positive perceptions are held regarding the *sub-goals*, whereas the *ultimate goals* are not being judged as positively. This is especially apparent in 74.1% of the respondents indicating that EA is a (very) good instrument to provide insight into the complexity of the organization (resulting in our acceptance of this hypothesis), whereas merely 29.4% of the respondents indicate that EA is a (very) good instrument to control the complexity of the organization (resulting in our rejecting this claim). Also, 71.9% state that EA is a (very) good instrument to depict a clear image of the desired future situation, and 55.6% that EA is a (very) good instrument to standardize, integrate and/or eliminate redundant processes and systems (resulting in the acceptance of both statements). However, a mere 13.4% actually indicates that EA is a (very) good instrument to control costs. The relationship with the outside world could also be better, with only 28.2% stating that EA is a (very) good instrument to co-operate with other organizations effectively and efficiently, and 25.3% that it is a (very) good instrument to react to changes in the outside world in an agile fashion. Also, EA does not seem to be a highly effective means for achieving business/IT alignment. In other words, EA yields several benefits that are valuable in their own right and conditional for obtaining further value, but, as yet, has not achieved its full promised potential.

Looking at Tables 7 and 8, we see that the number of nonrespondents is much higher for questions regarding projects than for those regarding the organization as a whole. The free-text option at the end of the survey provides some clues as to why. For example, some Enterprise Architectures focus not on internal project decisions, but on

| Table 7. Evaluations of EA benefits for projects | | | | | | | | |
|---|---|---|---|---|---|---|---|---|
| **H1 statement**<br>Projects conforming to EA turn out to… | **Much more often** | **More often** | **As often** | **Less often** | **Much less often** | **NR** | **P-value** | **H1?** |
| B12. …exceed their budgets less often than projects that do not have to conform to EA. | 2.4% | 15.7% | <u>60.8%</u> | 19.3% | 1.8% | 127 | 0.000* | ✘ |
| B13. … exceed their deadlines less often than projects that do not have to conform to EA. | 1.7% | 21.3% | <u>59.2%</u> | 16.7% | 1.1% | 119 | 0.000* | ✘ |
| B15. …deliver the desired quality more often than projects that do not have to conform to EA. | 3.7% | <u>53.4%</u> | 37.0% | 5.3% | 0.5% | 104 | 0.000 | ✔ |
| B16. …deliver the desired functionality more often than projects that do not have to conform to EA. | 3.8% | 40.4% | <u>45.4%</u> | 9.8% | 0.5% | 110 | 0.136 | ✘ |

e.g. high-level project portfolios instead. Some respondents also indicated that all projects had to conform to EA, making it hard to distinguish between conforming and non-conforming initiatives. Furthermore, working with EA was relatively new in several organizations, making it difficult to evaluate projects in the context of architecture, especially because these questions were quite detailed. Table 7 shows that projects conforming to EA clearly deliver higher quality. This is interesting, as we included this quality claim in the questionnaire for reasons of completeness, not because of its prevalence in literature (although claims that EA helps the organization to comply with quality standards such as ISO 9001 can be found). No such convincing support was found for delivering this type of project within time and budget limits. Rejection of better delivery of functionality is somewhat less convincing, since respondents did provide positive evaluations, albeit not in sufficiently convincing numbers. Furthermore, although it appears that EA simply does not offer projects much in the way of time and cost savings, it should also be noted that respondents have scored none of these four standard aspects very negatively.

| Table 8. Evaluations of EA benefits and conformance for projects | | | | | | | | |
|---|---|---|---|---|---|---|---|---|
| **H1 statement**<br>Projects that have to conform to EA turn out to… | **Much better** | **Better** | **Neither better nor worse** | **Worse** | **Much worse** | **NR** | **P-value** | **H1?** |
| B14. …be better equipped to deal with risks than projects that do not have to conform to EA. | 2.9% | <u>48.3%</u> | 43.1% | 5.2% | 0.5% | 82 | 0.001 | ✔ |
| B17. …be better equipped to deal with complexity (of the project and/or its immediate environment) than projects that do not have to conform to EA. | 5.3% | <u>63.6%</u> | 26.7% | 4.0% | 0.4% | 68 | 0.000 | ✔ |
| **H1 statement** | **Much slower** | **Slower** | **Neither slower nor faster** | **Faster** | **Much faster** | **NR** | **P-value** | **H1?** |
| B18. …get initialized faster than projects that do not have to conform to EA. | 5.4% | <u>45.9%</u> | 38.3% | 10.4% | 0.0% | 71 | 0.000* | ✘ |
| **H1 statement** | **Always** | **Frequently** | **Sometimes** | **Seldom** | **Never** | **NR** | **P-value** | **H1?** |
| O1. Projects that are required to conform to EA turn out to actually conform to the architectural principles, models and other prescriptions. | 4.2% | <u>61.0%</u> | 26.7% | 8.1% | 0.0% | 57 | 0.000 | ✔ |
| O2. Principles, models and other architectural prescriptions turn out to be open to multiple interpretations. | 2.7% | 27.8% | <u>57.0%</u> | 11.8% | 0.8% | 30 | 0.000 | ✔ |

Interestingly, whereas respondents reported that EA is not particularly capable of controlling complexity at the organizational level, EA does seem to enable projects to deal with complexity at the project level. Given the narrower scope of projects, this makes sense. Organization-wide ambitions will have to deal with far more incompatible systems, processes and stakeholder interests than more locally oriented projects. In addition, EA may provide compliant projects with an advantage over other projects by giving them insight into organizational complexity (for which, as we have seen, EA is very well-equipped). This enables projects to deal with complexity at the local level. A similar explanation may be provided for the fact that the results support the statement that projects conforming to EA are better equipped to deal with risks.

Also interesting is the fact that we did not find support for the hypothesis that projects conforming to EA are initialized faster than projects that do not have to conform to EA. In fact, significantly more than 40% (p=0.000) of the respondents stated that these projects actually start up (much) slower than projects not conforming to EA. This might be the result of the additional commitment that EA brings to bear on projects, such as getting acquainted with EA prescriptions, undergoing compliance assessments, dealing with additional stakeholders and balancing possible conflicts between local and enterprise-wide interests. Furthermore, the assumption behind many claimed project benefits, namely that many decisions have already been taken in the EA, may very well be questioned. In this respect, however, it is a notable finding that according to the respondents projects generally do tend to conform to EA.

The results also demonstrate that principles, models and other architectural norms turn out to be open to multiple interpretations. However, considering the results of Foorthuis et al. (2009), more respondents could have been expected to go for *Always* or *Frequently*. Most respondents opted for *Sometimes*, which is not the strongest support for the alternative hypothesis. One reason might be that prescriptions in most organizations are only moderately ambiguous. Alternatively, it could be that compliance assessments are usually carried out in a collaborative fashion, resulting in reduced disagreement (ibid.). People might also implicitly assume agreement, with disagreement only manifesting itself when being explicitly confronted with it.

We also asked why projects are being assessed on compliance with EA. Of the 275 respondents who stated that projects are being assessed (T4) *Seldom*, *Sometimes*, *Frequently* or *Always*, 49.5% indicated that this is being done in the context of obtaining permission from management for the actual implementation (a "building permit"). Furthermore, 34.5% indicated that it is input for a formal management decision (e.g. re-aligning), 49.5% that the reason is creating project awareness of deviations from the EA, and 28.7% that it is done for post-project awareness.

### Differences Between EA Creators and EA Users

We will now focus on the fourth sub-question. *Chi-square* tests (Norušis 2008) were employed to study whether statistically significant differences exist between evaluations of EA creators and EA users. The table below presents the percentage of EA creators and users giving *Good* or *Very Good* scores, the non-applicable respondents (NAR), the p-value of a $2 \cdot 2$ Pearson chi-square test with 1 degree of freedom, and whether we accept there is a difference in evaluation between EA creators and EA users. We excluded respondents who are simultaneously EA creator and EA user, which explains why the number of reported non-applicable respondents is conspicuously high (especially for projects, see the results in Table 10). The tables only show the differences that are either statistically significant or inconclusive.

| Table 9. EA creator and user evaluations of EA benefits for the organization as a whole | | | | | |
|---|---|---|---|---|---|
| **H1 statement**<br>EA turns out to be a good instrument to… | **EA creators:**<br>**(Very) Good** | **EA users:**<br>**(Very) Good** | **NAR** | **P-value** | **Difference?** |
| B2. …achieve an optimal fit between IT and the business processes it supports. | 51.6% | 37.9% | 95 | 0.052 | ? |
| B3. …provide insight into the complexity of the organization. | 78.6% | 66.3% | 86 | 0.048 | ✔ |
| B5. …integrate, standardize and/or deduplicate related processes and systems. | 64.6% | 47.6% | 91 | 0.015 | ✔ |
| B8. …co-operate with other organizations effectively and efficiently. | 32.6% | 18.6% | 118 | 0.034 | ✔ |
| B9. …depict a clear image of the desired future situation. | 81.0% | 66.3% | 84 | 0.017 | ✔ |
| B10. EA turns out to be a good frame of reference to enable different stakeholders to communicate with each other effectively. | 53.5% | 35.6% | 91 | 0.011 | ✔ |
| B11. EA, in general, turns out to be a good instrument. | 71.0% | 53.3% | 88 | 0.009 | ✔ |

As can be seen from Tables 9 and 10, in cases demonstrating a statistically significant difference, EA creators consistently prove to be more positive in their evaluation. This is demonstrated convincingly by the fact that creators are significantly more positive on the question of whether EA, in general, turns out to be a good instrument (B11). Although both groups are quite positive, 71.0% of the EA creators state that EA is a (very) good instrument, against 53.3% of the EA users. In addition, the two statements related to EA's communicative power (depicting a clear image and enabling effective communication) also show a difference. Other significant differences at the organizational level concern providing insight into complexity, standardization and integration. At the project level, delivering desired quality and functionality, and dealing with complexity and risks demonstrate significant differences.

| Table 10. EA creator and user evaluations of EA benefits for projects | | | | | |
|---|---|---|---|---|---|
| **Statement**<br>Projects conforming to EA turn out to… | **EA creators:**<br>**(Much) more often** | **EA users:**<br>**(Much) more often** | **NAR** | **P-value** | **Difference?** |
| B15. …deliver the desired quality more often than projects that do not have to conform to EA. | 75.4% | 42.3% | 157 | 0.000 | ✔ |
| B16. …deliver the desired functionality more often than projects that do not have to conform to EA. | 52.5% | 33.3% | 163 | 0.028 | ✔ |
| **Statement** | **(Much) less often** | **(Much) less often** | **NAR** | **P-value** | **Difference?** |
| B13. …exceed their deadlines less often than projects that do not have to conform to EA. | 24.1% | 10.6% | 169 | 0.045 | ✔ |

| Statement | (Much) better | (Much) better | NR | P-value | Difference? |
|---|---|---|---|---|---|
| B14. …be better equipped to deal with risks than projects that do not have to conform to EA. | 60.8% | 33.3% | 141 | 0.001 | ✔ |
| B17. …be better equipped to deal with complexity (of the project and/or its immediate environment) than projects that do not have to conform to EA. | 76.9% | 55.8% | 129 | 0.004 | ✔ |
| Statement | (Much) slower | (Much) slower | NR | P-value | Difference? |
| B18. …get initialized faster than projects that do not have to conform to EA. | 45.3% | 60.2% | 130 | 0.057 | ? |
| Statement | Always or Frequently | Always or Frequently | NR | P-value | Difference? |
| O2. Principles, models and other architectural prescriptions turn out to be open to multiple interpretations. | 20.8% | 39.8% | 99 | 0.004 | ✔ |

In short, EA creators are significantly more positive than EA users on several issues. Social psychology literature provides crucial insights for explaining these evaluative differences. Due to their involvement and commitment, EA makers should be regarded as subjective sources of information on EA. The binding effect of an earlier commitment (i.e. becoming an enterprise architect) results in them possessing a relatively positive attitude towards EA, and to not being easily persuaded by critical signals on its effectiveness (Zimbardo and Leippe 1991). EA users may be no less subjective, though, as they can not view the overall picture due to their local focus. Moreover, in their projects users have to deal with an additional effort when conforming to EA (as evidenced by the fact that all parties agree that these projects tend to get initialized more slowly). This may temper their enthusiasm regarding EA. Therefore, in order to have a balanced view, it is of paramount importance to take both perspectives into account.

## A Model for EA Conformance and Benefits

To investigate which *techniques* have a significant effect on *project conformance to EA* and identify the determinants of *EA benefits*, we carried out several analyses. In order to identify which techniques are associated with project conformance, we started out by measuring ordinal association (using Kendall's tau-b and Spearman's rho) and carried out univariate regression analyses. This revealed statistically significant bivariate associations between project conformance to EA (O1) and eight out of ten individual techniques (in descending order, according to tau-b: T4, T3, T7, T6, T2, T9, T5, T8). However, in order to control for the influence of other factors and explain the variation of dependent variable project conformance in terms of each technique's unique contribution, we needed to combine the variables in a single *multivariate regression model*. Since dependent variable conformance was measured on a five-point scale, we opted to employ ordinal regression[1] to construct the model (cf. Chen and Hughes 2004; Norušis 2009; Weisburd and Britt 2007). This logistic technique is based not on least squares, but on cumulative probabilities instead. In terms of assumptions, ordinal regression requires neither a normal distribution nor identical variance, but there is the strict demand of parallel slopes (which requires that the effects of the independent variables are constant across all categories of the dependent variable). Several link functions are available, each of which performs best in a specific condition. For example, the logit (odds ratio) is typically applied for evenly distributed categories, while the complementary log-log (cloglog) performs better when higher categories are more probable (Norušis 2009).

We subsequently built two full multivariate regression models, comprising all independent variables. One model included all techniques, the other included all techniques as well as all control variables. Due to a possible complete separation in the data (cf. So 1993) it could not be determined for these models whether the assumption of parallel slopes was violated or not. Moreover, because the number of cells explodes when many variables are included, dispersion of respondents results in poorer cell filling, making it harder to obtain statistically significant results. Despite the above, both full models did point our attention to three techniques that, even with all independent variables in the model, were highly statistically significant: management propagation of EA (T3), compliance assessments of projects (T4) and assistance for projects (T7). Since the other techniques and contextual variables are

---

[1]   Although some researchers use linear regression with 5-level ordinal variables (Garson 2010), we do not consider this the most appropriate technique here, especially given the fact that alternatives are widely available in software packages nowadays. Also note that we use the term "multivariate" to refer to a model with multiple independent variables (cf. Weisburd and Britt 2007).

**Parameter Estimates**

| | | Estimate | Std. Error | Wald | df | Sig. |
|---|---|---|---|---|---|---|
| Threshold | [ProjectsConform=2] | -6.299 | .803 | 61.480 | 1 | .000 |
| | [ProjectsConform=3] | -4.508 | .772 | 34.086 | 1 | .000 |
| | [ProjectsConform=4] | -1.061 | .572 | 3.439 | 1 | .064 |
| Location | [ComplianceAssessments=2] | -2.339 | .473 | 24.434 | 1 | .000 |
| | [ComplianceAssessments=3] | -2.513 | .442 | 32.347 | 1 | .000 |
| | [ComplianceAssessments=4] | -1.050 | .361 | 8.460 | 1 | .004 |
| | [ComplianceAssessments=5] | 0[a] | . | . | 0 | . |
| | [ManagementPropagation=2] | -1.521 | .521 | 8.505 | 1 | .004 |
| | [ManagementPropagation=3] | -.914 | .479 | 3.641 | 1 | .056 |
| | [ManagementPropagation=4] | -1.218 | .483 | 6.369 | 1 | .012 |
| | [ManagementPropagation=5] | 0[a] | . | . | 0 | . |
| | [Assistance=2] | -.951 | .479 | 3.943 | 1 | .047 |
| | [Assistance=3] | -.982 | .466 | 4.437 | 1 | .035 |
| | [Assistance=4] | -.903 | .436 | 4.298 | 1 | .038 |
| | [Assistance=5] | 0[a] | . | . | 0 | . |

a. This parameter is set to zero because it is redundant.

**Model Fitting Information**

| Model | -2 Log Likelihood | Chi-Square | df | Sig. |
|---|---|---|---|---|
| Intercept Only | 270.554 | | | |
| Final | 63.058 | 207.496 | 9 | .000 |

Link function: Complementary Log-log.

**Pseudo R-Square**

| | |
|---|---|
| Cox and Snell | .596 |
| Nagelkerke | .691 |
| McFadden | .457 |

Link function: Complementary Log-log.

controlled for, this is a very strong indication that these three techniques are important determinants of compliance. From here, we built various candidate models by combining several variables, from which it became clear that these three were indeed consistently the most robust techniques and were therefore included in the final model. Variables were recoded to obtain a similar meaning (i.e. a higher number always represents more benefits or more intensive use of a technique) and we collapsed the first two levels of each variable. The tables above present the SPSS output for the techniques and conformance section of the entire model (see Figure 2).

Note that in ordinal regression the coefficients are determined at the level of a variable category (although for continuous independent variables only one coefficient is calculated). Although at least one category of every independent variable should be statistically significant for the variable to be included in the model (Garson 2010), all of the concerning techniques have multiple categories significant at the 0.05 level. Furthermore, no variable has more than one category with a significance level above 0.05 and all categories are significant at the 0.075 level. In addition, the assumption of parallel slopes is not violated with a p-value of 0.472. The complementary log-log link function, which performs best when higher categories are more likely, provided the best results. The three techniques combined result in a good model fit, with a Nagelkerke $R^2$ of 0.691 (with that of the logit model being 0.427).

When interpreting the model, we see that all relationships are positive. Although counter intuitive, in ordinal regression a positive association is represented by negative coefficients. Interpret it as follows for any given technique: after controlling for the other two independent variables in the model, respondents in organizations in which the technique in question is never or seldom used (i.e. the collapsed level 2) are less likely to assign higher conformance scores than respondents in organizations in which the technique is always used (i.e. level 5, or the reference category). In short, the more a technique is used, the higher the achieved level of conformance. The Estimates (coefficients) provide clues as to the magnitude of the effects. Although the complementary log-log link function yields superior predictive and explanatory modeling power here, its coefficients are difficult to interpret (SPSS 2008). Also, as mentioned above, ordinal regression yields multiple coefficients for a factor when it is ordinal by nature. We therefore calculated a standardized regression coefficient for each independent variable in order to be able to compare the effects of the techniques on conformance. For this, we calculated the frequency-weighted standard error for each factor. While still difficult to interpret in an absolute sense, the resulting standardized coefficients indicate the relative effect size of the techniques. These coefficients are placed near the respective independent variables of Figure 2, while the Nagelkerke $R^2$ (which is easier to interpret) is placed near the dependent variable. From the standardized coefficients we can infer that being assessed on conformance (T4) has the most effect on whether projects will actually conform. The fact that a project will be explicitly confronted with its nonconformance apparently stimulates them to comply with the norms. This could be due to the fact that carrying out compliance assessments is an indication of the importance of conforming or, alternatively, simply to the project's desire to avoid confrontation. Management propagation of the importance of EA (T3) has the second largest influence. Third in rank is providing assistance to the projects in applying the EA's rules and guidelines (T7).

When considering the *benefits*, the results show that a statistically significant positive bivariate association and univariate regression coefficient exists between project conformance and whether EA in general is found to be a good instrument (B11). Running a multivariate regression analysis with all techniques plus conformance as independent variables to explain dependent variable B11 yields one variable with two significant categories (O1 conformance) and two variables with one significant category (T2 EA choices linked to business goals, and T4 compliance assessments). A similar argument can be made for the other benefits, where the strongest association is mostly with conformance rather than with the techniques. This supports the assertion that the three identified techniques indeed do not have a strong direct beneficial effect, but work via project conformance instead.

We subsequently studied the relationships between conformance and the individual benefits at both the project and organizational level. We started out by identifying statistically significant (p < 0.05) positive associations (using Kendall's tau-b and Spearman's rho). Next, we created separate univariate regression models, including the conformance and the benefit in question. Finally, we studied whether contextual variables could be included to enhance the results with a multivariate model. Regarding the benefits, the logit link function performed slightly better than the complementary log-log. The most interesting effects of project conformance on the benefits are depicted in Figure 2, again featuring frequency-weighted standardized coefficients for benefits that are explained by multiple determinants. A complete overview of statistically significant results can be found in Table 11.

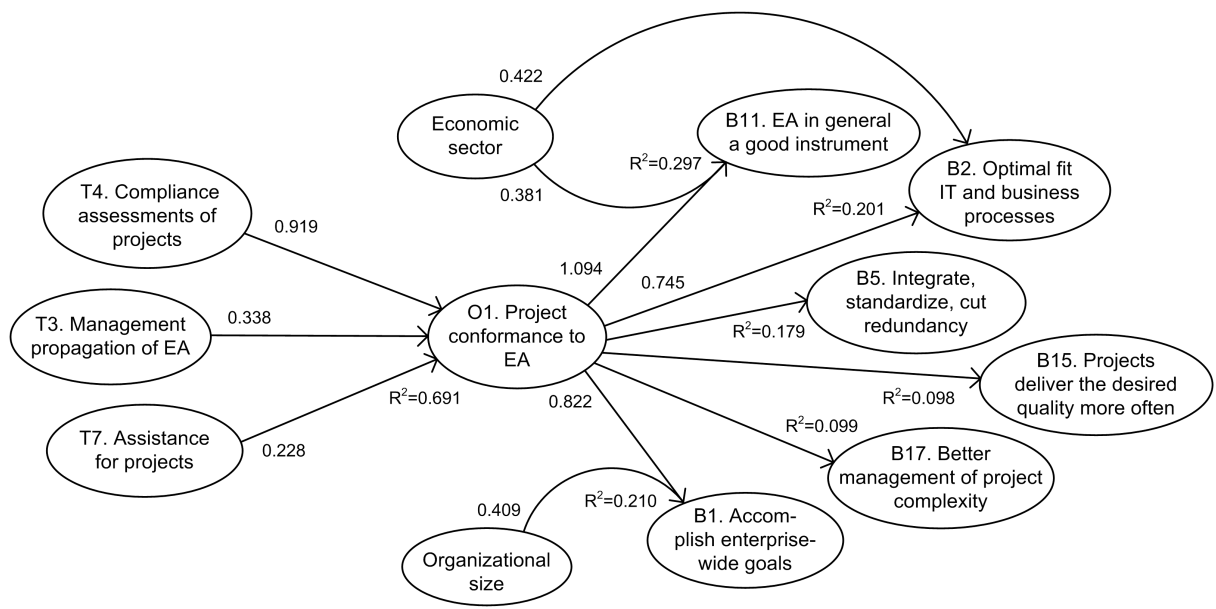

**Figure 2. The empirical model for EA conformance and benefits.**

Four *project benefits* could be attributed to conformance, namely delivering more of the desired functionality (B16) and quality (B15), and better management of complexity (B17) and risks (B14). If there is a high level of conformance, then projects conforming to EA – compared to non-conforming projects – are likely to achieve higher levels of these benefits. Quite interesting is that these results are highly consistent with the results of the direct answers in Tables 7 and 8, which demonstrated that only hypotheses B14, B15 and B17 were accepted. These findings are rather convincing: they show the internal validity of both the model (as conformance apparently plays a crucial role as a central variable) and the dataset (since direct responses and indirect associations are consistent, which definitely cannot be taken for granted because the respondents are largely unaware of the latter). Furthermore, although the delivery of desired functionality (B16) was rejected in Table 7, this benefit is also positively associated with project conformance. This indicates that it may be a benefit of which respondents are not yet fully aware. Indeed, re-examining the corresponding p-value in Table 7 reminds us of the fact that the functionality hypothesis was most definitely not rejected as convincingly as the other hypotheses. However, it should be pointed out that a positive association can in principle exist without ever achieving the two highest categories of the benefit. We leave it to the reader to draw his or her own conclusions.

Interestingly, although quite some claimed project benefits proved not to be significantly associated to project conformance, all *enterprise-wide benefits* were indeed significant. The strongest relationships were found for accomplishing enterprise-wide goals (B1), achieving an optimal fit between IT and business processes (B2) and integrating, standardizing and/or eliminating redundancy from related processes and systems (B5). These represent some of the key aims of Enterprise Architecture, the achievement of which in the respondents' experience is dependent on project compliance with EA. Especially for B2 and B5 the important role of projects is not difficult to see: business/IT alignment in processes and integrating several individual systems are typically EA-related issues, but the organization is highly dependent on projects for actual implementation. It is therefore not entirely unexpected to find strong relationships with project conformance here. Weaker, but still statistically significant associations were found for other important goals, such as achieving organizational agility (B7) and providing insight into and controlling the complexity of the organization (B3 and B4). Apparently, project conformance can contribute to resolving the complexity issue, but less so than to the other goals. Controlling costs (B6) and communicating the desired future situation and other concepts (B9 and B10) are somewhere in between in terms of being determined by conformance.

As with the project benefits, we observe the consistency between acceptance of simple hypotheses and the results of the regression analysis. The benefits with the highest Nagelkerke $R^2$s in Table 11 are accepted in Table 6 (B1, B5, B10, B11). We also see that B2 is barely insignificant in Table 6, which may again point to an unrecognized benefit.

| Table 11. Overview of significant effects on benefits | | | | | | | | |
|---|---|---|---|---|---|---|---|---|
| **Benefit associated with O1 Project Conformance to EA** | **Kendall's tau-b** | | **Univariate Ordinal Regression (Logit link function)** | | | **Multivariate Ordinal Regression (Logit link function)** | | |
| | **Value** | **P-value** | **Nagelkerke Pseudo $R^2$** | **Model p-value** | **Smallest p-value indep** | **Nagelkerke Pseudo $R^2$** | **Model p-value** | **Smallest p-values indep** |
| B11. EA, in general, is a good instrument | 0.424 | 0.000 | 0.268 | 0.000 | 0.000 ** | 0.297 | 0.000 | 0.000 ** |
| Economic sector (related to B11) | N.A. | N.A. | 0.040 | 0.006 | 0.002 ** | | | 0.000 |
| B1. Accomplish enterprise-wide goals | 0.353 | 0.000 | 0.166 | 0.000 | 0.000 * | 0.210 | 0.000 | 0.000 * |
| Organizational size (related to B1) | 0.059 | 0.266 | 0.028 | 0.069 | 0.032 | | | 0.000 |
| B2. Achieve optimal fit between IT and business processes | 0.333 | 0.000 | 0.159 | 0.000 | 0.000 ** | 0.201 | 0.000 | 0.000 ** |
| Economic sector (related to B2) | N.A. | N.A. | 0.055 | 0.001 | 0.000 ** | | | 0.001 |
| B5. Integrate, standardize and/or deduplicate related processes and systems | 0.336 | 0.000 | 0.179 | 0.000 | 0.000 * | | | |
| B10. Good frame of reference to enable different stakeholders to communicate with each other | 0.231 | 0.000 | 0.111 | 0.000 | 0.000 ** | | | |
| B6. Control costs | 0.270 | 0.000 | 0.104 | 0.000 | 0.007 | | | |
| B17. Better management of project complexity. | 0.199 | 0.002 | 0.099 | 0.000 | 0.001 | | | |
| B15. Projects deliver the desired quality more often | 0.230 | 0.001 | 0.098 | 0.002 | 0.013 | | | |
| B7. Respond to changes in the outside world in an agile fashion | 0.219 | 0.000 | 0.081 | 0.001 | 0.000 * | | | |
| B9. Depict a clear image of the desired future situation | 0.233 | 0.000 | 0.075 | 0.005 | 0.049 | | | |
| B16. Projects deliver the desired functionality more often | 0.209 | 0.003 | 0.070 | 0.012 | 0.007 | | | |
| B4. Control the complexity of the organization | 0.204 | 0.001 | 0.063 | 0.003 | 0.016 | | | |
| B3. Provide insight into the complexity of the organization | 0.197 | 0.001 | 0.060 | 0.003 | 0.002 | | | |
| B14. Better management of project risks | 0.157 | 0.018 | 0.059 | 0.017 | 0.003 | | | |
| B8. Co-operate with other organizations effectively & efficiently | 0.203 | 0.001 | 0.055 | 0.006 | 0.016 * | | | |

 * Multiple categories of the independent variable statistically significant at the 0.05 level
** All categories of the independent variable statistically significant at the 0.05 level

As can be seen from Figure 2 and Table 11, two contextual variables also make a contribution (visually modeled by arcs). The economic sector partly explains EA being a good instrument (B11), with the *public administration* sector being significantly outperformed by the *financial and insurance* and especially the *others* sectors. The same holds true for business/IT alignment (B2). It therefore seems that it is relatively difficult for public institutions to reap benefits from EA. A reason for this may be the fact that their individual organizational units often have high degrees of responsibility and autonomy, making it harder to demand compliance with standards that are suboptimal from a unit's point of view. Organizational size is a predictor for the extent to which enterprise-wide goals can be accomplished (B1), as organizations with up to 500 workers perform significantly better than organizations with at least 2,000 employees. On the other hand, although we consider the p-value of 0.083 inconclusive here, there are indications that organizations with 500 to 2,000 workers perform less well than organizations with more than 2,000 employees. As can be concluded from the size of the standardized coefficients, the contributions of the contextual factors are relatively small compared to those of project conformance. When *economic sector* is dropped from the model, conformance alone still yields an $R^2$ of 0.268 when explaining B11, and an $R^2$ of 0.159 when explaining B2. When *organizational size* is omitted, conformance alone still yields an $R^2$ of 0.166 when explaining B1. It should be noted that significant effects of *economic sector* were also observed for B7, B15 and B16. However, since the parallel slopes assumption was violated, these were not included in the model. The same holds true for contextual variables in other cases, making it a suitable topic for future research.

Another, more important, remark needs to be made regarding the magnitude of the effects. The Nagelkerke $R^2$s point to the fact that the effects of project conformance on enterprise-wide benefits are larger than the effects on more locally oriented project-level benefits. Regardless of whether contextual factors are included in the models, the benefits with the highest $R^2$s (i.e. above 0.100) are at the organizational level, e.g. B1, B2 and B5. This implies that (project conformance to) EA is indeed an important factor in achieving enterprise-wide benefits and goals. This is consistent with the accepted hypothesis B1 (see Table 6), which states that EA is a good instrument to accomplish enterprise-wide goals, instead of possibly conflicting local optimizations. As this is one of the key claims of Enterprise Architecture, these are important findings.

It was verified whether interaction effects exist between included factors, which could increase the explanatory power of the model, but none were found. Furthermore, all relationships were checked for confounding effects by controlling for the influence of contextual variables *economic sector*, *organizational size* and *EA focus*. However, the techniques (when explaining conformance) and conformance (when explaining benefits) retained their significant effects, regardless of whether or not the contextual variables themselves significantly contributed to the model.

Furthermore, since Figure 2 is actually comprised of several regression models (each Nagelkerke $R^2$ represents a separate model), the model as a whole does not have the property of transitivity. Therefore, effects of techniques cannot be calculated in terms of benefits. Take also into account that the project benefit variables are not absolute measures but represent the degree to which conforming projects outperform non-conforming projects on the respective aspects. Finally, the relatively low Nagelkerke $R^2$s on the right-most arrows do not mean that these benefits are not observed in practice (revisit Tables 6 to 8 to verify this), but that conformance and economic sector only explain part of the variation of the EA benefits. Further research should therefore focus on identifying other relevant factors, such as quality aspects of the EA itself.

## Conclusions and Further Research

This survey study has yielded several contributions to the field of EA. In answering the first research question, we have shown which techniques for stimulating compliance with EA are used in practice. The research results show that, except for sanctioning, all techniques identified are used in practice. To answer the second and third research questions, we presented the evaluative perceptions of people who have to deal with EA in a professional capacity. Evaluations prove to be positive on many accounts, both for individual systems development projects and the enterprise as a whole, but sub-goals (e.g. gaining insight into complexity) seem to be more easily achieved than ultimate goals (e.g. actually controlling complexity). With regard to the fourth research question, we have shown that EA creators and EA users differ in their evaluations regarding EA on many accounts, with the former having a relatively positive attitude. As an implication, future research on EA should take both perspectives into account, so as to prevent one-sided representation and less valid results. In answering the fifth research question, several positive relationships were found between techniques, conformance and benefits. We started out by identifying eight significant associations between compliance stimulating techniques and actual conformance. Using multivariate regression analysis, three techniques in particular have been identified which together explain project conformance to EA to a large extent. Finally, compliance with EA is shown to be positively associated with several benefits at both the project and the organizational level. These findings also show that project compliance is an important factor in obtaining value from the usage of EA. This is not only because the identified techniques seem to work *via* conformance, but also because conformance helps in realizing the key goal of gaining *enterprise-wide benefits*. The empirical results therefore establish project compliance with EA as a crucial factor in organizational performance. Another finding of this study is that some of EA's benefits may not yet be fully recognized by its practitioners.

Although our dataset and model demonstrated highly consistent results, there are several limitations to consider. First, we have been measuring perceptions of respondents instead of objective results. This is not problematic, however, as this is often the case with evaluative survey research. Moreover, perceptions have long ago been established as a valid indicator of organizational performance and so-called objective measures tend to have their own shortcomings (Dess and Robinson 1984; Venkatraman and Ramanujam 1987; Wall et al. 2004). Furthermore, our dataset proves to be quite internally valid, as direct responses (Tables 6, 7 and 8) and associations prove to be highly consistent on crucial aspects. Secondly, the usual limitations of causal analysis based on observational rather than experimental data apply. This also shows that the use of the aforementioned perceptions is a satisfactory approach, since using 'objective' measures (such as actual costs) will not yield more valid results in a non-laboratory setting. This is simply because it is impossible to fully control for non-EA factors, such as organizational culture when measuring simultaneously, or economic crises in the case of longitudinal research. That being said, however, regression analysis is an excellent method to simulate a true laboratory setting as much as non-experimental settings allow.

Thirdly, although the EA conformance and benefits model adheres to the strict assumptions of ordinal regression, it is important to mention its limitations. At 50%, the number of empty cells is high, rendering some additional chi-square Goodness-of-Fit measures unreliable (although they did pass). On the other hand, it does not seem that the empty cells are the result of biasing nonresponse. Empty cells are in part the result of the skewed distribution – which made binomial testing and the complementary log-log link function so effective – and the relative absence of extreme values in general. Consequently, this does not lead to the conclusion that a bias exists, as empty cells simply represent empirically less relevant categories. Reducing the number of empty cells can be achieved by collapsing adjacent levels. However, as ordinal information is lost, statistical power decreases (rendering levels statistically insignificant). It also often results in the parallel slopes test yielding an either violated or inconclusive outcome. Another drawback of using ordinal regression, or at least the complementary log-log, is the fact that interpreting the results is less precise (although this does prevent the false precision to which other methods are prone).

Despite the limitations mentioned, we consider this a valuable contribution nonetheless. Not only because it offers important insights into the factors that determine project compliance with and effectiveness of EA, but also because no other research seems to quantitatively model project conformance at all. Note that our insights may also be relevant for other forms of conformance, e.g. regulatory compliance. Furthermore, our findings are not only interesting from an academic perspective, but also highly relevant for practitioners.

However, because of the limitations and lack of comparative research, some modesty is appropriate. We therefore consider the model presented here to be a first version, as it should be tested further. In this context, we have several suggestions for future research. First, a more sophisticated concept of project conformance might yield more fine-grained insights. A distinction could be made between different aspects, for example architectural compliance regarding business, information, applications and infrastructure (cf. Boh and Yellin 2007). Alternatively, the four compliance aspects (or compliance checks) as described in Foorthuis et al. (2009) could be used to structure the conformance concept. Finally, a distinction could be made between compliance regarding project deliverables and the process of executing a project. Note that more sophisticated constructs may result in even higher levels of item-nonresponse than we observed, as the questions will be more detailed and therefore relatively difficult to answer. A second suggestion for future research concerns the fact that conformance and the contextual variables only explain part of the variation of the benefits. More factors in addition to or in interaction with conformance should therefore be taken into account, e.g. various EA practices and quality aspects, or the type or size of projects. Thirdly, additional analysis methods could be used in order to study the effects that remain hidden in this study due to the shortcomings of ordinal regression (such as the violation of the parallel slopes test). Regardless of the specifics of future research, however, this study has clearly shown that EA offers different kinds of value, but that additional effort is required from the IS community to fulfill more of its promised potential.

**Acknowledgements**. The authors wish to thank Abby Israëls, Priscilla Chandrasekaran, Pascal van Eck, Deirdre Giesen, Slinger Jansen, Nico Brand, Ilja Heitlager, Michael van Eck, Frank Hofman, Maarten Emons, Dick Kroeze, Rintske Thibault and the ICIS reviewers for their valuable remarks.

# References

Armour, F.J., Kaisler, S.H., Liu, S.Y. (1999). A Big-picture Look at Enterprise Architectures. *IT Professional* (1:1), pp. 35-42.

Bernus, P. 2003. "Enterprise models for enterprise architecture and ISO9000:2000". *Annual Reviews in Control* (27:2), pp. 211–220.

Boh, W.F., Yellin, D. 2007. "Using Enterprise Architecture Standards in Managing Information Technology". *Journal of Management Information Systems* (23:3), pp. 163-207.

Bucher, T., Fisher, R., Kurpjuweit, S., Winter, R. 2006. "Enterprise architecture Analysis and Application. An Exploratory Study". In: *Proceedings of TEAR 2006*, EDOC Workshop. URL: http://tear2006.telin.nl

Capgemini. 2007. "Enterprise, Business and IT Architecture and the Integrated Architecture Framework". URL: http://www.au.capgemini.com/m/en/tl/tl_Enterprise__Business_and_IT_Architecture_and_the_Integrated_Architecture_Framework.pdf

Chen, C.K., Hughes Jr., J. 2004. "Using Ordinal Regression Model to Analyze Student Satisfaction Questionnaires". *IR Applications* (1:1).

Dess, G.G., Robinson, R.B. 1984. "Measuring Organizational Performance in the Absence of Objective Measures: The Case of thePrivately-Held Firm and Conglomerate Business Unit". *Strategic Management Journal* (5:3), pp. 265-273.

Ellis, D., Barker, R., Potter, S., Pridgeon, C. 1993. "Information Audits, Communication Audits and Information Mapping: A Review and Survey". *International Journal of Information Management* (13:2), pp. 134-151.

Foorthuis, R.M., Brinkkemper, S. 2008. "Best Practices for Business and Systems Analysis in Projects Conforming to Enterprise Architecture". *Enterprise Modelling and Information Systems Architectures* (3:1), pp. 36-47.

Foorthuis, R.M., Hofman, F., Brinkkemper, S., Bos, R. 2009. "Assessing Business and IT Projects on Compliance with Enterprise Architecture". In: *Proceedings of GRCIS 2009*, CAISE Workshop on Governance, Risk and Compliance of Information Systems.

Garson, G.D. 2010. "Ordinal Regression (Statnotes from North Carolina State University)". URL: http://faculty.chass.ncsu.edu/garson/PA765/ordinalreg.htm

Gregor, S., Hart, D., Martin, N. (2007). "Enterprise Architectures: Enablers of Business Strategy and IS/IT Alignment in Government". *Information Technology & People* (20:2), pp. 96-120.

Goodhue, D.L, Kirsch, L.J., Quillard, J.A., Wybo, M.D. 1992. "Strategic Data Planning: Lessons from the Field". *MIS Quarterly* (16:1), pp. 11-34.

Henderson, J.C. & Venkatraman, N. 1993. "Strategic alignment: Leveraging information technology for transforming organizations" (Reprint). *IBM Systems Journal* (38:2/3), 1999, pp. 472-484.

Jonkers, H., Lankhorst, M.M., Doest, H.W.L. ter, Arbab, F., Bosma, H., & Wieringa, R.J. (2006). Enterprise Architecture: Management Tool and Blueprint for the Organisation. *Information Systems Frontiers* (8:2), pp. 63-66.

Kappelman, L., Pettite, A., McGinnis, T., Sidorova, A. 2008. "Enterprise Architecture: Charting the Territory for Academic Research". In: *Proceedings of AMCIS 2008*, Americas Conference on Information Systems.

Lankhorst, M. et al. 2005. Enterprise Architecture at Work. Modelling, Communication and Analysis. Berlin: Springer.

Morganwalp, J.M., Sage, A.P. 2004. "Enterprise architecture measures of effectiveness". *International Journal of Technology, Policy and Management* (4:1), pp. 81-94.

Mulholland, A., Macaulay, A.L. 2006. "Architecture and the integrated architecture framework". URL: http://architectes.capgemini.com/communauteDesArchitectes/laMethodologieIAF/b_Architecture_and_the_Integrated_Architecture_Framework.pdf

Niemi, E. (2006). Enterprise Architecture Benefits: Perceptions from Literature and Practice. In: *Proceedings of the 7th IBIMA Conference on Internet & Information Systems in the Digital Age*, Brescia, Italy.

Norušis, M.J. 2008. *SPSS Statistics 17.0 Guide to Data Analysis*. Upper Saddle River, NJ: Prentice Hall.

Norušis, M.J. 2009. *SPSS 17.0 Advanced Statistical Procedures Companion*. Upper Saddle River, NJ: Prentice Hall.

Obitz, T., Babu K., M. 2009. "Enterprise Architecture Expands its Role in Strategic Business Transformation". *Infosys Enterprise Architecture Survey 2008/2009*.

The Open Group. 2009. TOGAF Version 9: The Open Group Architecture Framework. TOG.

Pulkkinen, M., Hirvonen, A. 2005. "EA Planning, Development and Management Process for Agile Enterprise Development". In: *Proceedings of the 38th Hawaii International Conference on System Sciences*, pp. 223.3.

Raadt, B. van der, Soetendal, J., Perdeck, M., Vliet, H. van. (2004). Polyphony in Architecture. In: *Proceedings of the 26th International Conference on Software Engineering* (ICSE'04).

Richardson, G.L., Jackson, B.M., Dickson, G.W. 1990. "A Principles-Based Enterprise Architecture: Lessons from Texaco and Star Enterprise". *MIS Quarterly* (14:4), pp. 385-403.

Ross, J.W., Weill, P., Robertson, D. 2006. *Enterprise Architecture as Strategy: Creating a Foundation for Business Execution*. Boston, Massachusetts: Harvard Business School Press.

Slot, R., Dedene, G., Maes, R. 2009. "Business Value of Solution Architecture". In: E. Proper, F. Harmsen, J.L.G. Dietz (eds.). *Advances in Enterprise Engineering II*, LNBIP (28), pp. 84-108. Berlin: Springer.

So, Y. 1993. "A tutorial on logistic regression". In: *Proceedings of the 18th SAS Users Group International Conference*, 1290-1295.

SPSS. 2008. *SPSS 16.0 Help File*. SPSS Inc.

Van Steenbergen, M., Brinkkemper, S. 2009. "The architectural dilemma: division of work versus knowledge integration". In: H. Weigand, H. Werthner, G. Gal (eds.), *Proceedings of the Third International Workshop on Business/IT Alignment and Interoperability* (BUSITAL'09) held in conjunction with CAiSE'09 Conference, pp. 46-60, Amsterdam, the Netherlands.

Van Steenbergen, M., Schipper, J., Bos, R., Brinkkemper, S. 2010. "The Dynamic Architecture Maturity Matrix: Instrument Analysis and Refinement". In: A. Dan, F. Gittler, and F. Toumani (Eds.), *ICSOC/ServiceWave 2009, LNCS 6275*, pp. 48-61. Berlin: Springer-Verlag.

Venkatraman, N., Ramanujam, V. 1987. "Measurement of Business Economic Performance: An Examination of Method Convergence". *Journal of Management* (13:1), pp. 109-122.

Versteeg, G., Bouwman, H. 2006. "Business architecture: A new paradigm to relate business strategy to ICT". *Information Systems Frontiers* (8:2), pp. 91-102.

Wagter, R., Berg, M. van den, Luijpers, J., Steenbergen, M. van. 2005. *Dynamic enterprise architecture: How to Make It Work*. Hoboken, New Jersey: John Wiley & Sons.

Wall, T.D., Michie, J., Patterson, M., Wood, S.J., Sheehan, M., Glegg, C.W., West, M. 2004. "On the Validity of Subjective Measures of Company Performance". Personnel Psychology (57:1), pp. 95-118.

Ward, J., Peppard, J. 2002. *Strategic Planning for Information Systems*. 3rd Edition. Chichester: John Wiley & Sons Ltd.

Weisburd, D., Britt, C. 2007. "Multivariate Regression with Multiple Category Nominal or Ordinal Measures: Extending the Basic Logistic Regression Model". In: *Statistics in Criminal Justice*, 3rd Edition, pp. 579-606. New York: Springer.

Zimbardo, P.G., Leippe, M.R. 1991. *The Psychology of Attitude Change and Social Influence*. New York: McGraw-Hill.